\documentclass[journal=cmatex,manuscript=article]{achemso}
\usepackage[version=3]{mhchem} 
\usepackage{color}

\author{C\'{e}sar Moreno}
\email{cesar.moreno@icn2.cat}
\affiliation{Catalan Institute of Nanoscience and Nanotechnology (ICN2), CSIC and The Barcelona Institute of Science and Technology, Campus UAB, Bellaterra, 08193 Barcelona, Spain}
\author{Mirco Panighel}
\affiliation{Catalan Institute of Nanoscience and Nanotechnology (ICN2), CSIC and The Barcelona Institute of Science and Technology, Campus UAB, Bellaterra, 08193 Barcelona, Spain}

\author{Manuel Vilas-Varela}
\affiliation{Centro de Investigaci\'{o}n en Qu\'{i}mica Biol\'{o}xica e Materiais Moleculares (CIQUS) and Departamento de Qu\'{i}mica Org\'{a}nica, Universidade de Santiago de Compostela. Santiago de Compostela 15782, Spain}

\author{Guillaume Sauthier}
\affiliation{Catalan Institute of Nanoscience and Nanotechnology (ICN2), CSIC and The Barcelona Institute of Science and Technology, Campus UAB, Bellaterra, 08193 Barcelona, Spain}

\author{Gustavo Ceballos}
\affiliation{Catalan Institute of Nanoscience and Nanotechnology (ICN2), CSIC and The Barcelona Institute of Science and Technology, Campus UAB, Bellaterra, 08193 Barcelona, Spain}

\author{Diego Pe\~na}
\email{diego.pena@usc.es}
\affiliation{Centro de Investigaci\'{o}n en Qu\'{i}mica Biol\'{o}xica e Materiais Moleculares (CIQUS) and Departamento de Qu\'{i}mica Org\'{a}nica, Universidade de Santiago de Compostela. Santiago de Compostela 15782, Spain}

\author{Aitor Mugarza}
\email{aitor.mugarza@icn2.cat}
\affiliation{Catalan Institute of Nanoscience and Nanotechnology (ICN2), CSIC and The Barcelona Institute of Science and Technology, Campus UAB, Bellaterra, 08193 Barcelona, Spain}
\alsoaffiliation{ICREA – Instituci\'{o} Catalana de Recerca i Estudis Avan\c{c}ats, Lluis Companys 23, 08010 Barcelona, Spain}

\title
  {Critical role of phenyl substitution and catalytic substrate on the surface-assisted polymerization of dibromobianthracene derivatives}

\abbreviations{GNR, ARPES, XPS, STM}
\begin{document}

\begin{abstract}

Understanding the nature and hierarchy of on surface reactions is a major challenge for designing coordination and covalent nanostructures by means of multistep synthetic routes. In particular, intermediates and final products are hard to predict since reaction paths and their activation windows depend on the choice of both the molecular precursor design and the substrate. Here we report a systematic study of the effect of the catalytic metal surface to reveal how a single precursor can give rise to very distinct polymers that range from coordination and covalent non planar polymer chains of distinct chirality, to atomically precise graphene nanoribbons and nanoporous graphene. Our precursor consists on adding two phenyl substituents to 10,10’-dibromo 9,9’-bianthracene, a well-studied precursor in the on-surface synthesis of graphene nanoribbons. The critical role of the monomer design on the reaction paths is inferred from the fact that the phenyl substitution leads to very distinct products in each one of the studied metallic substrates.

\end{abstract}

\section{Introduction}

In the past few years, the on-surface synthesis of carbon-based nanoarchitectures out of single building-block precursors has attracted great attention \cite{Franc2011,Fan2015,Robert2015,Talirz2016,Nacci2016,Shen2017}. Recent advances in this active field are the synthesis of atomically precise graphene nanoribbbons (GNR) \cite{Cai2010,Ruffieux2016}, and covalent \cite{Bieri2009,Lafferentz2012,Lipton-Duffin2009,Basagni2015,Moreno2018} and metal-organic coordination \cite{Park2011,Saywell2014,Dong2016} polymeric chains and networks. One of the most common strategies for GNR synthesis has been the use of non-planar, halogen-based hydrocarbons in a two-step reaction: i) dehalogenative homocoupling polymerization (Ullmann coupling), and ii) formation of planar aromatic skeletons by cyclodehydrogenation \cite{Cai2010,Bjork2011,Talirz2016}. This strategy has led to GNRs with armchair, zigzag, and more complex edge structures \cite{Kimouche2015,Liu2015,Liu2016a,Ruffieux2016,Talirz2016,Oteyza2016}, to their atomically controlled functionalization \cite{Kawai2015,Nguyen2016,Carbonell-Sanroma2017a}, or to GNR heterostructures \cite{Chen2015,Dienel2015,Carbonell-Sanroma2017}.

Despite the above successful examples, predicting the final product from a given precursor and catalytic surface is far out of reach of present understanding. This is due to the subtle correlation of parameters such as the monomer structure and adsorption configuration \cite{Jacobse2016,Schulz2017}, the surface crystal structure and reactivity \cite{Oteyza2016,Simonov2018}, or the presence of byproducts  \cite{Bjork2013,Batra2014} and metal adatoms \cite{Simonov2018,Patera2017,Schulz2017}. An illustrative case of the critical role of the interplay between monomer and surface properties on the final product is that of dibromo bianthracene (DBBA) precursors. $10,10'$-Dibromo $9,9'$-bianthracene undergoes different reaction paths depending on the catalytic substrate.  Whereas the conventional Ullmann coupling route leads to straight armchair GNRs in Au(111) \cite{Cai2010,Simonov2014,Talirz2016} and Au(110) \cite{Massimi2015}, dehydrogenative cross coupling polymerization on Cu(111) leads to chiral zigzag edge GNRs \cite{Han2014,Han2015,SanchezSanchez2016}. On Ag(111), armchair GNRs are obtained through a very different path where polymerization occurs only after turning monomers into graphene platelets by a simoultaneous dehalogenation and cyclodehydrogenation \cite{Huang2012}. This covalent coupling of individual platelets can be, however, blocked by the strong interaction with the substrate, as found for Cu(110) \cite{Simonov2015}. Finally, replacing the Br sites from the $10,10'$ to the $2,2'$ sites minimizes the effect of the surface, univocaly guiding the reaction path through the Ullman coupling route in Au(111), Ag(111), and Cu(111) \cite{Oteyza2016}. Understanding the role of the monomer and the substrate in hierarchical on-surface reactions is therefore determinant to synthesize atomically precise GNRs, the main activity in this field, but also to design new synthetic routes for novel polymeric structures.

Here we present a systematic study of the thermally assisted reactions and the resulting structures obtained with the same monomer on Au(111), Ag(111), Ag(100), and Cu(111). For our studies we use the home-synthesized $2,2'$-diphenyl $10,10'$-dibromo $9,9'$-bianthracene (DP-DBBA), which can be seen as a derivative of $10,10'$-dibromo $9,9'$-bianthracene (DBBA), the most studied precursor for GNR synthesis. The temperature dependent X-ray (XPS) and ultraviolet (UPS) photoemission, and scanning tunneling microscopy (STM) experiments carried out on the different substrates gives us insight on the role of the substrate, whereas comparison with the available literature on the on-surface reactions of DBBA on each of the above substrates enables to discriminate pure monomer effects. The obtained final product is different for each substrate, and each of them differs also from that obtained with DBBA on the same substrate. The variety of the products we obtain range from coordination and covalent polymer chains where the precursor arrangement dictates their chiral imprint, to graphene nanoribbons and to nanoporous graphene.

\section{Results and discussion}

\subsection{The molecular precursor}
The precursor monomer used in this work, $2,2'$-diphenyl $10,10'$-dibromo $9,9'$-bianthracene (DP-DBBA), is a derivative of the well-studied $10,10'$-dibromo $9,9'$-bianthracene (DBBA) in the synthesis of graphene nanoribbons\cite{Cai2010, Ruffieux2016}, with one additional phenyl group added at opposite sides of each anthracene branch. The molecule is synthesized synthesized in a two-steps procedure starting from $2,2'$-dibromo-$9,9'$-bianthracene (1, Fig. 1a). First, double Pd-catalyzed Suzuki coupling of compound 1 with two equivalents of phenyl boronic acid affords diphenyl derivative 2. Then, careful dibromination leads to the isolation of compound DP-DBBA in good yield (see Supplementary Materials of Ref. \cite{Moreno2018} for details in the synthesis). 

\begin{figure}[ht!]
\begin{center}
			\includegraphics[width=8.5 cm]{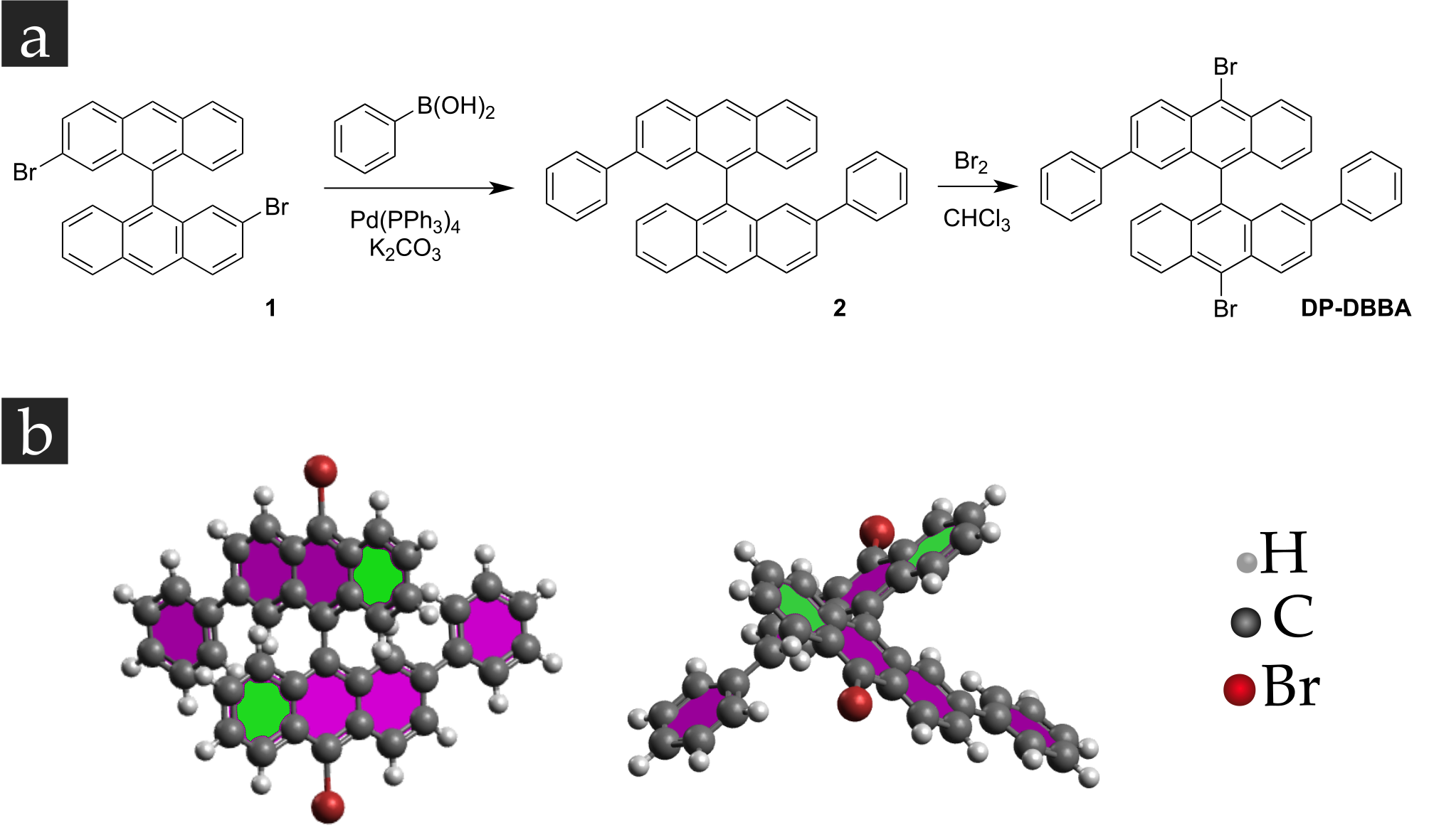}
	\caption{Synthesis and structure of monomer DP-DBBA. (a) Synthetic route to obtain $2,2'$-diphenyl $10,10'$-dibromo $9,9'$-bianthracene (DP-DBBA). (b) Top and side views of the gas-phase relaxed structure of DP-DBBA, where the staggered conformation arising from the steric hindrance between hydrogen atoms can be clearly visualized (the high-end aromatic rings are highlighted in green).}
	\label{Fig1}
	\end{center}
\end{figure}

Bianthryl derivatives are characterized by the staggered geometry induced by the steric repulsion between hydrogen atoms, which twists the anthracene subunits around the C-C bond connecting them, as can be seen in the schematic model of Fig. \ref{Fig1}b. Interestingly, the phenyl substituents imprint chirality on the DP-DBBA molecule, which can transferred to the polymer chains upon Ullmann coupling depending on the chiral mixture of precursor monomers, as will be shown latter.

\subsection{Au(111): From covalent polymeric chains, to nanoribbons and nanoporous graphene}

\begin{figure}[ht!]
\begin{center}
			\includegraphics[width=17.5 cm]{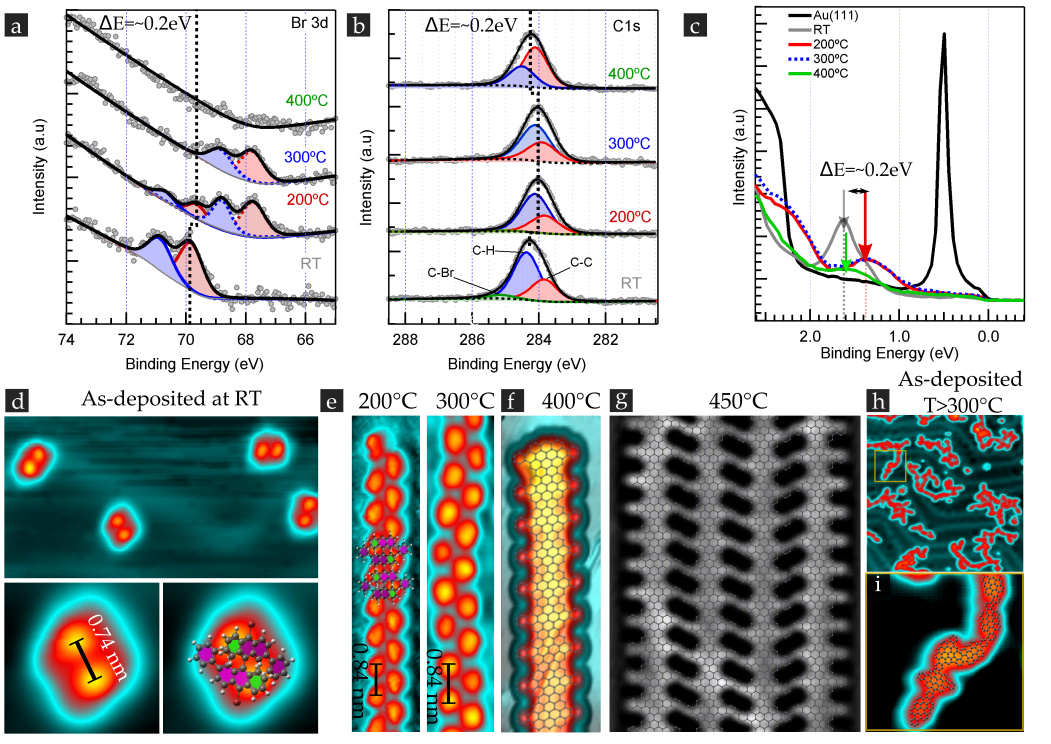}
	\caption{\scriptsize \textbf {Generation of polymeric chains, graphene nanoribbons and nanoporous graphene on Au(111).} \textbf{a}, Br 3d and \textbf{b} C 1s core-level spectra and \textbf{c}, UV photoemission spectroscopy of as-deposited monomer precursor at RT and progressive stepwise anneal stages.  Spectra fitting of C 1s considering C-C, C-H and C-Br deconvoluted components. Raw data is displayed by gray dots and the the fitting results and subcomponents by solid lines. \textbf{d}, Constant current STM topographic images of as-deposited precursor molecules: overview of 14x7 nm$^{2}$ (top) and zoomed region of 2.5x2.5 nm$^{2}$ (down) with the single monomer structure template overlaid (It=0.1nA, Vs= 0.7V), \textbf{e} polymeric chains after mild annealing at 200$^{\circ}$C and 300$^{\circ}$C with the polymer model overlaid and the side view at the bottom (7x1.5 nm$^{2}$,It=0.1nA, Vs= 0.7V), \textbf{f} single graphene nanoribbon after 400$^{\circ}$C (9x3 nm$^{2}$,It=0.1nA, Vs= 0.7V), \textbf{g}, Laplace filtered constant-height STM image of nanoporous graphene formation after further annealing at 450$^{\circ}$C and its overlapped structural model [80x80 nm$^{2}$, I$_{t}$=0.1nA, V$_{s}$=1.0V],  \textbf{h} overview of as-deposited monomers holding the sample over 300$^{\circ}$C (30x30 nm$^{2}$,It=0.7nA, Vs= 0.2V) and \textbf{i} zoom in  \textbf{h} highlighting the linking of several planar monomer units (6x6 nm$^{2}$). All images were recorded at 5K.}
	\label{Fig2}
	\end{center}
\end{figure}

Au(111) is the most studied surface for the on-surface synthesis of covalent nanostructures, in particular GNRs \cite{Talirz2016}. The latter is mainly due to the fact that the majority of halogenated hydrocarbon precursors undergo stepwise Ullmann coupling and cyclodehydrogenation reactions with clearly separated thermal windows. The new monomer employed in this study allows for a third hierarchical reaction step that couples adjancent GNRs via dehydrogenative cross coupling to give rise to a nanoporous graphene, as we have recently reported.\cite{Moreno2018}

We use temperature-dependent XPS and UPS spectra to track the thermal reactions that give rise to the different intermediate and final structures. Debromination of the precursors can be followed by tracking energy shifts of the Br $3d$ core level, whereas the spectral decomposition of the C $1s$ multiplet can be used to follow cyclodehydrogenation by measuring the ratio between C-C and C-H bond contributions, and the formation of C-metal coordination bonds.
Figures~\ref{Fig2}a and b show XPS spectra of the Br $3d$ and C $1s$ core levels and corresponding fits for different temperatures. 

The spectrum recorded after deposition at room temperature (RT) at the Br $3d$ region shows the spin-orbit split $3d_{5/2}-3d_{3/2}$ doublet at 70.9 and 69.9 eV, which can be assigned to C-bonded Br,\cite{DiGiovannantonio2013,Gutzler2014,Massimi2015,Batra2014,Basagni2015,Oteyza2016}indicating that the monomer is still brominated upon deposition. When the sample is annealed at 200$^{\circ}$C a significantly quenched Br-C doublet coexists with a more intense doublet that appears at 2.1 eV lower binding energy (BE), which can be assigned to Br-Au bond formation. \cite{DiGiovannantonio2013,Gutzler2014,Massimi2015,Batra2014,Basagni2015,Oteyza2016}.  This transition is attributed to the partial cleavage of C-Br bonds and the passivation of Br radicals by surface Au atoms. At 300$^{\circ}$C, the sole presence of the Br-Au doublet indicates that the monomers are completely debrominated. Finally the absence of any peak 400$^{\circ}$C indicate that  Br adatoms are desorbed from the surface in this last temperature window. 

In the C $1s$ region, the small energy differences between the C-C, C-H, and C-Br bond contributions result in a single broad peak that is difficult to deconvolute by fitting, forcing us to reduce the degrees of freedom by assuming some relative ratios\cite{note1}. At RT, a nice fit is obtained by assuming relative intensities of the pristine monomer stoichiometry (C-H:C-C:C-Br=26:12:2). The best fit at 200$^{\circ}$C and 300$^{\circ}$C is obtained by maintaining the relative weight of C-H bonds and converting the ratio of cleaved C-Br obtained from the Br $3d$ analysis into C-C bonds. This reveals two important points. First, that C radicals formed by the Br cleavage are saturated by C-C coupling, as expected for the Ullmann polymerization. Second, that cyclodehydrogenation still did not take place yet. At 400$^{\circ}$C, however, the C-H:C-C bond ratio has to be modified to the one corresponding to GNRs (C-H:C-C=12:24), a clear fingerprint of the cyclodehydrogenative aromatization. Further annealing at 450$^{\circ}$C not alter the C1s (not shown), and then the edge dehydrogenation that takes place to create the nanoporous graphene is out of our XPS sensitivity.


The effect of the intramolecular transformations described above on the frontier orbitals can be tracked by UPS, as shown in Fig. \ref{Fig2}c. Upon deposition at RT, we can identify a prominent molecular orbital peak at 1.6 eV. The peak broadens after each reaction step identified in the XPS analysis, which is in agreement with a gradual delocalization of molecular orbitals expected by the polymerization and aromatization.

Comparing all XPS and UPS data displayed in Figs. \ref{Fig2}a-c together, we notice a coincident shift of $\sim0.2$ eV of all peaks in the thermal regime of 200-400$^{\circ}$C. The overall shift of the spectra is a signature of a non-local effect such as a work function change, which can be attributed to the presence of Br adatoms in this temperature range.\cite{Pham2016,Oteyza2016} Br is indeed known to substantially modify the work function of gold. \cite{Bertel1980}

We combine our spectroscopic analysis with real space structural information provided by scanning tunneling microscopy (STM). Monomers deposited with the sample held at room temperature are observed as double bright protrusions in STM images, as shown in Fig.~\ref{Fig2}d. They display an apparent height of about 0.23 nm and a center-to-center distance of 0.74 nm. Considering the gas-phase relaxed structure displayed in Fig.~\ref{Fig1}, we can relate the measured inter-lobe distance with the two opposite, high end benzene rings of the scissor-like staggered anthracenes, and hence conclude that the monomer binds to the surface from the functional aryl group. Annealing at around 200$^{\circ}$C promotes the formation of 1D chains observed as a zigzag of bright protrusions with an apparent height of about 0.31 nm and a periodicity of 0.84 nm. This values are in excellent matching with the periodicity of C-C bonded DP-DBBA of 0.85 nm depicted in the schematics of Fig.~\ref{Fig2}e, and also with previously reported values for DBBA \cite{Cai2010}. The zigzag correlation confirms that the protrusions correspond to the non-functionalized side of the anthracene, since the presence of the aryl group at the high end would lead to the alignment of the protrusions pairs perpendicular to the chain. Further annealing at above 400$^{\circ}$C leads to protrusionless 1D chain structures with a reduced apparent height of 0.18 nm (see Fig. \ref{Fig2}f). The planarization of the backbone is in agreement with the aromatization of the polymers upon cyclodehydrogenation and consequent formation of GNRs \cite{Bjork2011}. Similar values were founded for other type of on-surface synthesized GNRs \cite{Cai2010}. Interestingly, the cyclization of the functional aryl group in our precursor leads to an atomic structure that differs substantially from the more conventional armchair GNRs obtained using plain DBBA, giving rise to particular periodic cove-shape edges which allow the formation of nanoporous graphene (Fig.~\ref{Fig2}g) after a subsequent annealing at 450$^{\circ}$C.\cite{Moreno2018}

The hierarchical reactions that lead to the synthesis of straight GNRs with well defined edge structures as the one shown in Fig. \ref{Fig2}f require clearly separated thermal windows for the dehalogenation and cyclodehydrogenation steps. Mixing the two steps by, for instance, rising the dehalogenation temperature by replacing Br with Cl leads to a disordered polymerization of individual graphene platelets formed by the aromatization of single precursors \cite{Jacobse2016} Here we show that the two steps can be kinetically mixed by modifying the annealing method. Long, straight GNRs are obtained by slowly annealing a precursor covered sample to 400$^{\circ}$C within 1 hour. In contrast, precursor deposition with the sample held at 300$^{\circ}$C leads to disordered polymer chains very similar to those obtained with chlorinated DCBA precursors in sequential annealing steps. The chains, as shown in Figs. \ref{Fig2}h and i, seem to be formed by randomly linked planar units of the size of the monomers, with a lateral size of 1.5 nm along the long axis direction, and a height of 0.15 nm.

\subsection{Ag(111): From racemic to chiral metal coordinated polymers}

GNRs have been synthesized on Ag(111) using DBBA \cite{Huang2012,Oteyza2016} and tetraphenyl-triphenylene \cite{Cai2010} precursors. Although a priori the results could be expected due to the similarity with Au(111) in lattice structure and chemical reactivity, a comparison of the intermediate structures obtained by DBBA reveal different chemical paths followed in each substrate. On Ag(111), the precursor is already debrominated and dehydrogenated at 180$^{\circ}$C, but instead of forming C-C coupled polymer chains as in Au(111), they self-assemble in a hexagonal pattern of flattened graphene platelets.\cite{Huang2012} The platelets polymerize into GNRs by increasing the annealing temperature to 380$^{\circ}$C. The results presented in the following illustrates how modifying the precursor from DBBA to DP-DBBA can further alter this path, giving rise to well-defined polymeric intermediates but inhibiting the formation of GNRs.

\begin{figure}[ht!]
\begin{center}
			\includegraphics[width=17.5 cm]{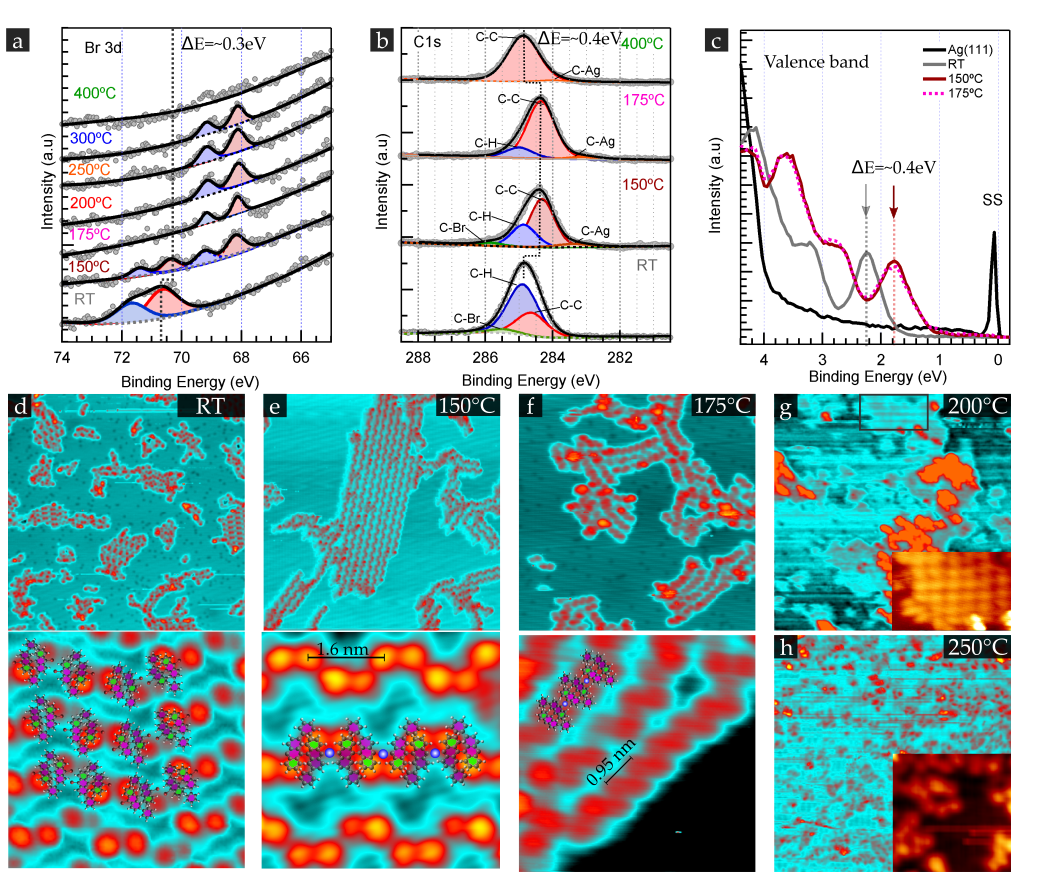}
	\caption{\scriptsize \textbf {Temperature-assisted on-surface reactions on Ag(111).} \textbf{a}, Br 3d and \textbf{b} C 1s core-level spectra and \textbf{c}, UV photoemission spectroscopy of as-deposited monomer precursor at RT and progressive stepwise anneal stages.  Spectra fitting of C 1s considering C-C, C-H and C-Br deconvoluted components. Raw data is displayed by gray dots and the the fitting results and subcomponents by solid lines. The subtracted background is marked by dotted lines. Constant current STM topographic images of as-deposited precursor molecules: \textbf{d}, overview of 56x56 nm$^{2}$ (top) and zoomed region of 5.6x5.6 nm$^{2}$ (down) (It=0.1nA, Vs=0.8V) , \textbf{e} annealed at 150$^{\circ}$C: overview of 56x56 nm$^{2}$ (top) and zoomed region of 5x5 nm$^{2}$ (down) (It=0.1nA, Vs=1.0V), \textbf{f} annealed at 175$^{\circ}$C (34x34 nm$^{2}$ (top) and zoomed region of 8.5x8.5 nm$^{2}$ (down) (It=0.1nA, Vs= 1.2V), \textbf{g} annealed at 200$^{\circ}$C (45x45 nm$^{2}$, inset:10.5x10.5 nm$^{2}$; It=0.1nA, Vs= 1.2V)  and, \textbf{h} annealed at 250$^{\circ}$C (63x63 nm$^{2}$, inset:15x15 nm$^{2}$; It=0.1nA, Vs= 0.7V). All images were recorded at 78K.}
	\label{Fig3}
	\end{center}
\end{figure}

Figures ~\ref{Fig3}a and b shows the evolution of Br $3d$ and C $1s$ core levels as a function of annealing temperature. As in Au(111), we find that the precursor adsorbs brominated at RT. This is in agreement with the intact adsorption of DBBA reported previously in \cite{Shen2017a} where the bromine atoms are placed in the same $10,10'$ site. However, we note that moving the Br to the  $2,2'$ site has shown to lower the debromination down to RT, reflecting a strong sensitivity of the dissociation barrier to structural details of the molecule on this substrate.\cite{Oteyza2016} At 150$^{\circ}$C, the coexistence of both Br-C and Br-Ag doublets separated by 2.2 eV reveals the onset of debromination, which is already completed at 175$^{\circ}$C. This lower dehalogenation temperature as compared to Au has also been found for DBBA.\cite{Huang2012} Finally, Br adsorbates desorb between 300$^{\circ}$C and 400$^{\circ}$C.  

C $1s$ core level fitting (Fig.~\ref{Fig3}b) confirms the intact adsorption of as-deposited monomers at RT.  A good fitting is obtained by constraining the relative C-Br, C-H and C-C peak intensities to the precursor stoichiometry. After annealing at 150$^{\circ}$C, however, the behaviour is different as in Au(111). Here the fitting requires an additional component at the lower BE tail, which is a signature of Br-Ag bonds.\cite{Gutzler2014, Basagni2015,Massimi2015,Simonov2015,Smerieri2016, Pis2016} Comparing the similar branching ratio of C-Br/C-Ag and Br-C/Br-Ag bonds we can conclude that the debrominated C radicals are passivated by metal coordination instead of covalently coupled to other molecules, as found in Au(111). On the other hand, a reduced intensity of the C-H component obtained at this temperature set an onset of partial dehydrogenation that is gradually completed at 400$^{\circ}$C (all series not shown). This is in contrast to DBBA, where dehydrogenation is already completed at 180$^{\circ}$C.\cite{Huang2012}  In the thermal range of 150-300$^{\circ}$C we also observe subtle peak shifts that can be related to the structural transformations induced by the gradual dehydrogenation. Finally the broad dominant C-C peak observed above 200$^{\circ}$C suggests that some kind of disordered covalent structures with different C-C bonding configuration are being formed at the high temperature regime (spectra at 400$^{\circ}$C shown as example).

Valence band spectra displayed in Fig. \ref{Fig3}c exhibits a clear molecular orbital at around 2.2 eV binding energy, 0.6 eV higher than in Au(111), which correlates to the work function difference between the two materials. In contrast to that found in Au(111), we do not observe any significant broadening of the molecular orbital level up to 175$^{\circ}$C. The absence of any signature of orbital delocalization is in line with the inhibition of covalent polymerization suggested by the C $1s$ peak fitting.

Similar to the case of Au, we also observe a general shift of all XPS and UPS peaks of 0.4 eV to lower binding energies in the thermal window where Br adatoms are present in the surface, attributed again to changes in the work function.

The two dimensional self-assembled structures observed by STM after deposition at room temperature differ from the arrays of parallel armchairlike chains found for DBBA (Fig. \ref{Fig3}d).\cite{Shen2017a} Since molecules adsorb intact at this temperature, we assume that the assembly is driven by weak, Br mediated interactions.\cite{Chung2011,Fan2014a,Fan2015a,Kawai2015a,Morchutt2015,Huang2016}. It is only after annealing to 150$^{\circ}$C when the 2D clusters transform into similar armchairlike chains. However, the precursor arrangement within the chain cannot be the one proposed for DBBA,\cite{Shen2017a} since that would overlap the functional phenyl groups of neighbouring molecules. Instead, we propose an alternative structure that consists on Ag coordinated chains (see Fig. \ref{Fig3}e), which would be in turn consistent with the detection of the C-Ag component with XPS. We note that this new arrangement also consists of alternating molecules with opposite chiral conformation, as the one proposed for DBBA. Interestingly, the racemic chains seem to transform into chiral chains after annealing to 200$^{\circ}$C, as shown in Fig. \ref{Fig3}f. This chiral phase separation, probably triggered by the different selectivity in the interactions of the partially dehydrogenated molecules, implies considerable molecular reorganization. The STM appearance of the chiral chains emulate the covalently coupled polymers found in Au(111), with molecules aligned with the anthracenes perpendicular to the chain (see Fig. \ref{Fig2}e). The measured periodicity of 0.95 nm, however, is 0.1 nm larger than the C-C coupled polymers, difference that can be attributed to a metal coordination. \cite{Oteyza2016} This is again supported by the sizeable C-Ag component found in the corresponding C $1s$ XPS spectrum, and the fact that  metal coordinated intermediates seem to be rather ubiquitous in the Ullmann coupling on Ag(111). \cite{Park2011,Gutzler2014,Oteyza2016,Dong2016} Finally, annealing to higher temperatures leads to disordered 2D and 1D polymerization instead of well-defined GNRs obtained with DBBA, as can be seen in Figs. \ref{Fig3}g and h.

\subsection{Ag(100): From non-planar to aromatic metal coordinated polymers}

\begin{figure}[ht!]
\begin{center}
			\includegraphics[width=17.5 cm]{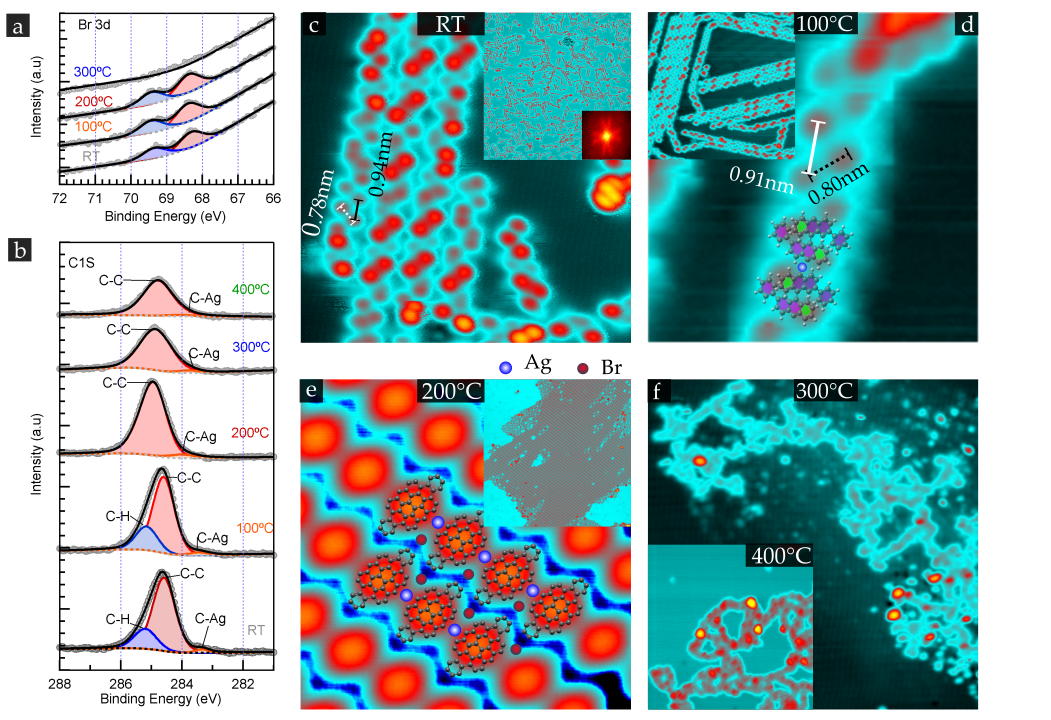}
	\caption{\scriptsize \textbf {Temperature-assisted on-surface reactions on Ag(100).} \textbf{a}, Br 3d and \textbf{b} C 1s core-level spectra of as-deposited monomer precursor at RT and progressive stepwise anneal stages. Spectra fitting of C 1s considering C-C, C-H and C-Br deconvoluted components. Raw data is displayed by gray dots and the the fitting results and subcomponents by solid lines. The subtracted background is marked by dotted lines. Constant current STM topographic images of: \textbf{c}, as-deposited precursor molecules  overview of 140x140 nm$^{2}$ (inset) and zoomed region of 14x14 nm$^{2}$ (It=0.1nA, Vs=1.0V) , \textbf{d} annealed at 100$^{\circ}$C: overview of 33.6x33.6 nm$^{2}$ (inset) and zoomed region of 5.6x5.6 nm$^{2}$ with the model proposed overlaid (It=0.1nA, Vs=1.2V), \textbf{e} annealed at 200$^{\circ}$C (84.4x84.4 nm$^{2}$ (inset) and zoomed region of 14x14 nm$^{2}$ (It=0.10nA, Vs= 1.0V) with the metal organic model proposed overlaid, \textbf{f} annealed at 300$^{\circ}$C (42x42 nm$^{2}$, It=0.1nA, Vs= 1.2V)  and (inset) annealed at 400$^{\circ}$C (42x42 nm$^{2}$, It=0.1nA, Vs= 0.4V). All images were recorded at 78K.}
	\label{Fig4}
	\end{center}
\end{figure}

On-surface reactions on Ag(100) have been studied very recently using the prototypical DBBA as precursor.\cite{Smalley2017} In this study, although long-range self-assembled molecular structures are observed at RT, annealing to 200$^{\circ}$C already results in the desorption of most of the material, leaving a few undefined clusters on the surface. It is only by direct deposition at 400$^{\circ}$C that polymerization gives rise to irregular chains that the authors assign to disordered GNRs. In contrast to these results, the higher stability we find for DP-DBBA on this substrate leads to well-defined polymeric structures to temperatures up to 200$^{\circ}$C, as shown in figure\ref{Fig4}.

XPS spectra obtained after deposition at RT reveal a lower debromination barrier on Ag(100) as compared to Ag(111) and Au(111). The energy of the Br $3d$ doublet shown in Fig. \ref{Fig4}a has the energy of the metal-bonded component, indicating that the molecule debrominates upon adsorption at RT. The higher temperature sequence shows that Br is stable on the surface up to T = 200$^{\circ}$C. Debromination below RT has also been observed 
for DBBA on Cu(110)\cite{Simonov2015} and Cu(111),\cite{Simonov2014} but our observation is in contrast to that reported for Ag(100),\cite{Smalley2017} where the authors concluded that DBBA adsorbed intact based on the absence of Br atoms in STM images. 

The C $1s$ spectrum obtained at RT can only be fitted by reducing the C-H/C-C ratio well below the values of the intact precursor and adding a C-Ag component, suggesting that both debromination and partial dehydrogenation has taken place simultaneously, and that C radicals are saturated by metal coordination (see Fig. \ref{Fig4}b). The bond ratio is maintained at 100$^{\circ}$C, but by increasing the annealing temperature to 200$^{\circ}$C the molecule undergoes complete dehydrogenation, according to the disappearance of the C-H component. The C-Ag component maintains roughly constant in intensity. Annealing to higher temperatures only leads to a broadening of the C-C peak and a decrease of the total intensity, signatures of partial desorption and bond disorder. 

The core level shift observed between 100$^{\circ}$C and 200$^{\circ}$C cannot be related to Br induced work function modifications, since similar amount of Br adatoms are found for the two temperatures. Alternatively, we can attribute the shift to a stronger interaction with the underlying metal of the planar aromatic structures obtained from the dehydrogenation, as revealed by the STM images discussed below. The more subtle shifts observed above this temperature are, on the other hand, an indication of further structural transformations occurring in this temperature range.

The 1D chains imaged by STM after deposition indicate that polymeric chains are already formed at RT, as shown in Fig. \ref{Fig4}c. The position of the protrusion pairs follows that of the chiral arrangement of the C-C coupled polymers on Au(111) and the metal coordinated ones found at 200$^{\circ}$C on Ag(111). The periodicity of 0.94 nm that we find points towards metal coordination, in agreement with the C $1s$ peak analysis (following the same arguments as for the Ag(111) case). We note that the protrusion pairs can be classified in two different heights of 0.22 and 0.31 nm, which we attribute to the partial dehydrogenation found by XPS. The structure of the partially dehydrogenated molecule, however, remains unsolved. We note here that, although the reduction of C-H bonds found in the C $1s$ spectra could imply the coexistence of fully dehydrogenated planar, and hydrogenated non-planar species, we discard this scenario since all species exhibit the double protrusion appearance characteristic of the staggered anthracene units. Planar oval-shaped structures are observed, however, at 200$^{\circ}$C, where XPS data show no sizeable contribution of C-H bonds (see Fig. \ref{Fig4}b). The measured dimensions of 1.0 nm and 1.4 nm along the short and long axis directions respectively are in good agreement with the size of the graphene platelet that would result from the cyclodehydrogenation of a single precursor. The measured apparent height of 0.15 nm is also very similar to that of GNRs and the disordered chains of graphene platelets  formed on Au(111) (Figs. \ref{Fig2}f-h). The planar units form large 2D islands that resemble arrays of parallel chains (inset Fig. \ref{Fig4}b). From the intrachain periodicity of 1.06 nm, and the remaining C-Ag contribution in the C $1s$ peak, we conclude that the graphene platelets within the chain are also linked by metal coordination. The interchain distance of 1.42 nm is very similar to those found for other self-assembled structures formed by DBBA-based graphene platelets, where the interactions seem to be mediated by intercalated Br adatoms.\cite{Huang2012,Simonov2015} Based on that, and supported by the fact that the adatoms we observe outside the clusters cannot account for all the released Br, we propose the structural configuration superimposed to the STM image in Fig. \ref{Fig4}e, where Br atoms are intercalated within the Ag-bonded chains. Indeed, a very similar configuration has been proposed for arrays of Ag-coordinated dibromo anthracene chains formed in Ag(111).\cite{Park2011} Heating the sample to higher temperatures (300-400$^{\circ}$C) leads to the fragmentation of the coordinated polymers and the desorption of roughly half of the organic material (Fig. ~\ref{Fig4}f and inset), in good agreement with the C1s core level broadening and  quenching.

\subsection{Cu(111): Disordered polymerization}

The polymerization mechanism followed by the DBBA precursor on Cu(111) leads to chiral zigzag GNR structures instead the armchair ones obtained in Au(111) and Ag(111). \cite{Han2014,SanchezSanchez2016} In this substrate, the C radicals resulting from dehalogenation are not relevant in the polymerization process, and instead neighbouring molecules link via dehydrogenative cross coupling of their low lying phenyls. The onset for the formation of the chiral GNRs is around 250$^{\circ}$C,\cite{Han2014,SanchezSanchez2016} and they are stable up to at least 500$^{\circ}$C.\cite{Han2014,SanchezSanchez2016} Our studies reveal a lower stability for DP-DBBA on this surface, as we show in the following.

\begin{figure}[ht!]
\begin{center}
			\includegraphics[width=17.5 cm]{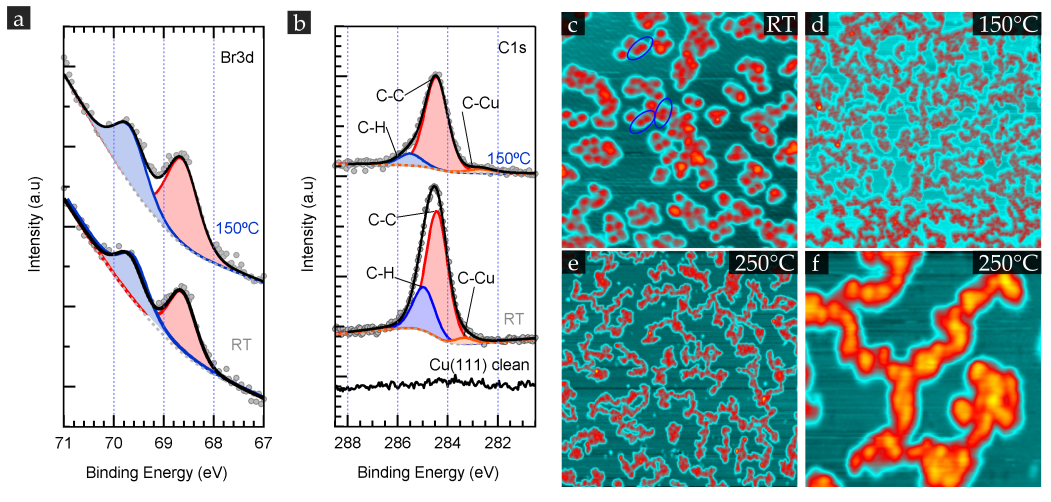}
	\caption{\scriptsize \textbf {Temperature-assisted on-surface reactions on on Cu(111).} \textbf{a}, Br 3d and \textbf{b} C 1s core-level spectra of as-deposited monomer precursor at RT and progressive stepwise anneal stages. Constant current STM topographic images of \textbf{c}, as-deposited precursor monomers of 17x17 nm$^{2}$, \textbf{d} annealed at 150$^{\circ}$C (56x56 nm$^{2}$), \textbf{e} annealed at 250$^{\circ}$C (56x56 nm$^{2}$ and \textbf{f} zoomed region of 12.5x12.5 nm$^{2}$.  All images were recorded at It=0.1nA, Vs= 1.0V and 78K.}
	\label{Fig5}
	\end{center}
\end{figure}

Similar to the case of Ag(100), monomers are debrominated upon deposition on Cu(111) at RT (Fig.~\ref{Fig5}a), as indicated by the position of the Br $3d$ doublet at 68.6 and 69.6 eV. Br adatoms remain on the surface after annealing to 150$^{\circ}$C, where we stopped the series based on the highly disordered structures already observed by STM (Fig.~\ref{Fig5}d). The C $1s$ fit depicts an scenario where the monomers are already substantially dehydrogenated and metal coordinated at RT (Fig.~\ref{Fig5}b). The dehydrogenation process continues at 150$^{\circ}$C, where, in addition, almost half of the species are already desorbed.

STM images obtained at RT present a disordered surface with apparently random structures. We find few protrusion pairs with the characteristic apparent height and protrusion pair separation of the monomer of 0.25-0.30 nm and 0.74 nm respectively (circles in Fig. \ref{Fig5}c). Yet, the distance between most protrusions exceed the molecular size, and their width of about 1 nm indicate that each of them is an individual reacted monomer. Annealing to 150$^{\circ}$C and 250$^{\circ}$C leads to  substantial desorption and a gradual formation of disordered chains that could be attributed to an ill-defined polymerization.

\section{Summary}

We have carried out a systematic, multitechnique study of the on-surface reactions and resulting structures obtained by using DP-DBBA, a newly synthesized bianthryl derivative as precursor, in comparison with previous studies using the well-known DBBA monomer.

Hierarchical Ullmann coupling and cyclodehydrogenation in Au(111) leads to the formation of well-defined straight GNRs of cove-shaped edges when the precursor is deposited at RT and annealed at low rates. Further annealing leads to an additional dehydrogenative step where GNRs couple laterally and form nanoporous graphene. Deposition over 300$^{\circ}$C however, results in the mixing of both reaction steps and randomly oriented covalent linking of fully aromatized monomers that form disordered GNRs.   

On Ag(111) a simultaneous dehalogenation and dehydrogenation sets at 150$^{\circ}$C. Instead of Ullmann polymerization, the C radicals are linked by Ag coordination, giving rise to metal-organic polymeric chain structures where chirality evolves with temperature. The inhibition of covalent C-C coupling results in an inferior stability of the coordinated chains as compare to the GNRs obtained in Au(111). 

Metal coordinated polymer chains are already formed at RT on the more reactive Ag(100). A different thermal evolution as compared to Ag(111) leads to a full aromatization of the monomer units at 200$^{\circ}$C that form large 2D islands of arrays of Ag-coordinated chains. The lack of covalent linking limits the stability of these structures below 300$^{\circ}$C.

Finally, on Cu(111) we do not observe any ordered polymerization, but rather irregular chains of likely covalently bonded species.

The radically different structures obtained on each surface, and the comparison to similar studies on DBBA illustrate the difficulties on predicting on-surface reactions and related structures. Whereas DBBA leads to well-defined (albeit different type of) GNRs on Au(111), Ag(111) and Cu(111), DP-DBBA only forms GNRs on Au(111). Instead, the latter gives rise to metal coordinated chains on the Ag substrates. Conversely, the lack of any ordered structure obtained with DBBA on Ag(100) is in contrast with the long-range ordered polymeric chains obtained with DP-DBBA. Altogether, our study highlights the decisive role of both design of the monomer and choice of the substrate in the on-surface synthesis of covalent nanoarchitectures.

\section{Experimental methods}

The experiments were performed in two separate UHV systems, one dedicated to STM, the other to high resolution core level XPS and UPS. Both UHV systems (base pressure $<5\times 10^{-10}$ mbar) were equipped with an ion sputter gun for surface cleaning.
Sample preparation was carried out identically in both systems. Clean metal surfaces was obtained by cycles of $Ar^{+}$ bombardment ($<$1 keV) at RT and annealing at 400-450$^{\circ}$C for 5-10 min in ultra-high vacuum. Surface cleanliness was confirmed with x-ray photoelectron spectroscopy (XPS) prior monomer deposition. Monomer deposition were sublimated at 304$^{\circ}$C comprising times between 3 and 10 min for coverages until the monoloayer.  XPS measurements were carried out using a Specs Phoibos 150 hemispherical energy analyser using a monochromatic X-ray source (Al K$_{\alpha}$ line with an energy of 1486.6 eV and 400 W) and energy referenced to the Fermi level. UPS experiments were performed with incident light from the He I emission at 21.2 eV. The fits have been performed with Voigt functions (Lorentzian-Gaussian curves), with a Gaussian-Lorentzian ratio of 0.3. Fitting was performed using XPST macro for IGOR (Dr. Martin Schmid, Philipps University Marburg) using the minimum number of peaks required to minimize the R-factor. Temperatures were measured from a infrared pyrometer at the XPS setup, which was previously calibrated by fixing a thermocouple directly to the sample holder. In the STM setup, the temperature was and directly measured with a thermocouple spot-welded to the sample holder during the annealings. STM images were processed with WsXM software \cite{Horcas2007}.For the synthesis of DP-DBBA from 2,2,’-dibromo-9,9’-bianthracene see Supplementary Materials of Ref. \cite{Moreno2018}.

\begin{acknowledgement}
C.M was supported by the Agency for Management of University and Research grants (AGAUR) of the Catalan government through the FP7 framework program of the European Commission under Marie Curie COFUND action 600385. Funded by the CERCA Programme / Generalitat de Catalunya. ICN2 is supported by the Severo Ochoa program from Spanish MINECO  (Grant No. SEV-2013-0295). We acknowledge support from the Ministerio de Ciencia e Innovaci\'{o}n (MAT2013-46593-C6-2-P, MAT2013-46593-C6-5-P, MAT2016-78293-C6-2-R, MAT2016-78293-C6-4-R), Ag\'{e}ncia de Gesti\'{o} d\'{}Ajuts Universitaris i de Recerca-AGAUR (2014 SGR715), the EU project PAMS (610446)  the Xunta de Galicia (Centro singular de investigacion de Galicia accreditation 2016-2019, ED431G/09), and the European Regional Development Fund.

\end{acknowledgement}


\providecommand{\latin}[1]{#1}
\makeatletter
\providecommand{\doi}
  {\begingroup\let\do\@makeother\dospecials
  \catcode`\{=1 \catcode`\}=2 \doi@aux}
\providecommand{\doi@aux}[1]{\endgroup\texttt{#1}}
\makeatother
\providecommand*\mcitethebibliography{\thebibliography}
\csname @ifundefined\endcsname{endmcitethebibliography}
  {\let\endmcitethebibliography\endthebibliography}{}


\begin{tocentry}

Some journals require a graphical entry for the Table of Contents.
This should be laid out ``print ready'' so that the sizing of the
text is correct.

Inside the \texttt{tocentry} environment, the font used is Helvetica
8\,pt, as required by \emph{Journal of the American Chemical
Society}.

The surrounding frame is 9\,cm by 3.5\,cm, which is the maximum
permitted for  \emph{Journal of the American Chemical Society}
graphical table of content entries. The box will not resize if the
content is too big: instead it will overflow the edge of the box.

This box and the associated title will always be printed on a
separate page at the end of the document.

\end{tocentry}

\end{document}